\documentclass[a4paper]{jpconf}
\usepackage{graphicx}
\newcommand{\beqn}{\begin{equation}}
\newcommand{\eeqn}{\end{equation}}

\renewcommand{\>}{\rangle}

\begin{document}
\title{\boldmath $D^0 - \bar D^0$ mixing: theory basics}

\author{Diego Guadagnoli}

\address{Physik-Department, Technische Universit\"at M\"unchen, D-85748 Garching, Germany}

\ead{diego.guadagnoli@ph.tum.de}

\begin{abstract}
I discuss how the novel experimental data on $D^0 - \bar D^0$ mixing can be combined to provide 
information on the fundamental theoretical quantities describing the mixing itself.
I then discuss the theoretical impact of the new data, focusing in particular on the MSSM.
\end{abstract}

\noindent For times much longer than the strong interaction time scale, the flavor eigenstates 
$D^0 = (c \bar u)$ and $\bar D^0$ mix into each other. Mixing is a purely quantum effect and
the $D^0 - \bar D^0$ system is the only one featuring it among `up-type-quark' mesons, since 
the top quark decays before forming a bound state with an antiquark. The time evolution
producing the mixing is calculated through 
\beqn
i \frac{d}{dt}\left( \begin{array}{c} |D^0(t)\> \\ |\bar D^0(t)\> \end{array} \right) 
= \left( \hat M - \frac{i}{2} \hat \Gamma \right)\left( \begin{array}{c} |D^0(t)\> \\ 
|\bar D^0(t)\> \end{array} \right), 
~~\mbox{with}~~
\hat M = \left( \begin{array}{cc} M & M_{12} \\ M_{12}^* & M \end{array} \right),~~
\hat \Gamma = 
\left( \begin{array}{cc} \Gamma & \Gamma_{12} \\ \Gamma_{12}^* & \Gamma \end{array} \right),
\label{Ham}
\eeqn
where the explicit form for the hermitian $\hat M$ and $\hat \Gamma$ matrices holds assuming
CPT invariance. Mass eigenstates, with masses $m_{1,2}$ and widths $\Gamma_{1,2}$, are
defined through
\beqn
| D_{1,2} \> = p | D^0 \> \pm q | \bar D^0 \>~,~~\mbox{with}~\left( \frac{q}{p} \right)^2 =
\frac{M_{12}^* - \frac{i}{2} \Gamma_{12}^*}{M_{12} - \frac{i}{2} \Gamma_{12}}~,
\label{D12}
\eeqn
allowing in turn to define the basic mixing observables as $x = (m_2 - m_1)/\Gamma$ and 
$y = (\Gamma_2 - \Gamma_1)/(2\Gamma)$ \cite{revs}. If $|q/p| = 1$ in eq. (\ref{D12}), then
$|D_{1(2)}\>$ is CP even (odd), since one can choose phases so that $| D^0 \>
\stackrel{\rm CP}{\leftrightarrow} | \bar D^0 \>$ (see Y.Nir in \cite{revs}). If, on the other 
hand, $|q/p| \ne 1$, then mass eigenstates cannot be chosen as CP eigenstates and there is 
CP violation in mixing.

Within the Standard Model (SM), meson mixings are well described, in the $B_{d,s}$ and $K$
cases, by box diagrams with loop-exchange of $W$-bosons and up-type quarks. The flavor
structure of the contributions is the product of a CKM factor and an Inami-Lim function
$S_0(m_{q_1}^2,m_{q_2}^2)$, summed over the quark flavors $q_1, q_2$ running in the loop.
If $m_{q_1}^2, m_{q_2}^2 \ll M_W^2$, as is the case for all the down-type quarks, one has e.g.
$S_0(m_{q_1}^2,m_{q_1}^2) \simeq m_{q_1}^2 / M_W^2$, showing a very effective GIM
suppression. One should also note that in the $D$-mixing case, the third family
contribution, which would be enhanced by the relatively large $m_b$ mass, is accidentally
suppressed by a very small CKM factor, resulting in a relative box contribution from the
third family of O($10^{-3}$), and a correspondingly suppressed amount of CP violation 
within the SM. This is also in sharp contrast with the $K, B_{d,s}$ cases, where the third 
family (top) contribution is always important or dominant. Due to the above reasons, for
$D$-mesons SM (box) contributions are tiny, in principle making the mixing ideal room for 
New Physics to show up. On the other hand, the charm mass is accidentally of the same order of the
hadronic scale. Hence, $K, \pi$ intermediate states are likely to dominate the mixing
amplitude contributions. Generic predictions are $x_{\rm box} \le 10^{-5}$
and $x_{\rm long~dist.} \le O(10^{-3})$ (Burdman \& Shipsey in \cite{revs},\cite{petrov}).
The poor control over the long distance contributions presently impairs an effective use
of $D$-mixing as a test of the SM.

A channel that very simply illustrates how to experimentally access $D$-mixing is that of
``wrong sign'' $D \to K \pi$ decays. The amplitude for the decay $D^0 \to K^+ \pi^-$ proceeds 
in fact through the sum of a tree-diagram, which is however doubly Cabibbo-suppressed
(DCS) and indicated henceforth as $\mathcal D_{\rm DCS} \propto \sin^2 \theta_C$, and a diagram 
in which the $D^0$ oscillates first into a $\bar D^0$ whose Cabibbo-favored (CF) final state 
is then exactly $K^+ \pi^-$. The latter diagram behaves as $\mathcal D_{\rm mix + CF} \propto \cos^2 \theta_C$ 
but is suppressed by the loop factor of the mixing, which is what one wants to access. 
Hence the two diagrams are competitive and the mixing measurable \footnote{One should note
that in practice the term most easily allowing access to the mixing variables is the
interference between $\mathcal D_{\rm DCS}$ and $\mathcal D_{\rm mix + CF}$.}.

A special comment deserves CP violation in the $D$-system. A reasonable assumption is to
consider CP violation in decay amplitudes (`direct' CP violation) negligible, since the latter 
are dominated by the tree-level CP conserving SM contributions. On the other hand, non-negligible 
CP violation can occur in the mixing amplitude, due to non-SM short-distance contributions. 
In the case of the wrong sign $D \to K \pi$ decays, CP violation in mixing should however
be hard to observe, while likely to be accessible is CP violation in the {\em interference} 
between the decay with ($\mathcal D_{\rm mix + CF}$) and without ($\mathcal D_{\rm DCS}$) 
mixing \cite{revs}. The latter is related to the phase $\phi = \arg(q/p)$. Recalling the
definition of $q/p$, eq. (\ref{D12}), and parameterizing $M_{12} = |M_{12}|
\exp(-i \Phi_{12})$, $\Gamma_{12} = |\Gamma_{12}|$, with $\Phi_{12}$ small, one easily
recognizes that $\phi \approx + \Phi_{12} \times 4 |M_{12}|^2/(4 |M_{12}|^2 +
|\Gamma_{12}|^2)$. Thereafter, a naive estimate of $\Phi_{12}$ from the SM box contributions to 
the mixing gives $\Phi_{12} \le 10^{-2}$. Observation of (large) CP violation would then be a clear NP
signature, immune to hadronic uncertainties. For a recent critical analysis on this issue,
see Ref. \cite{ball}.

\begin{figure}[htb!]
\vspace{-0.5cm}
\begin{center}
\includegraphics[scale=0.3]{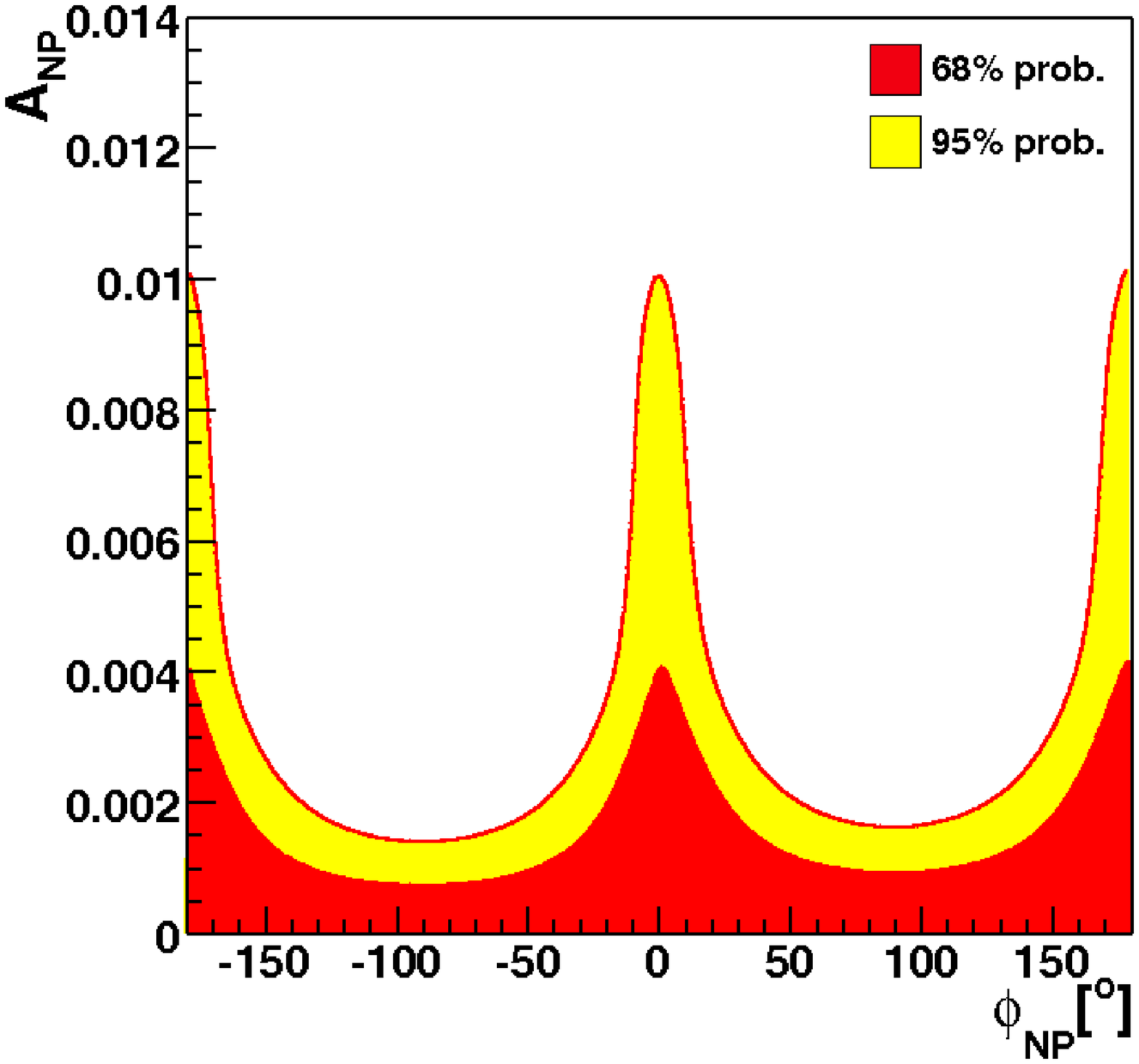}
\includegraphics[scale=0.3]{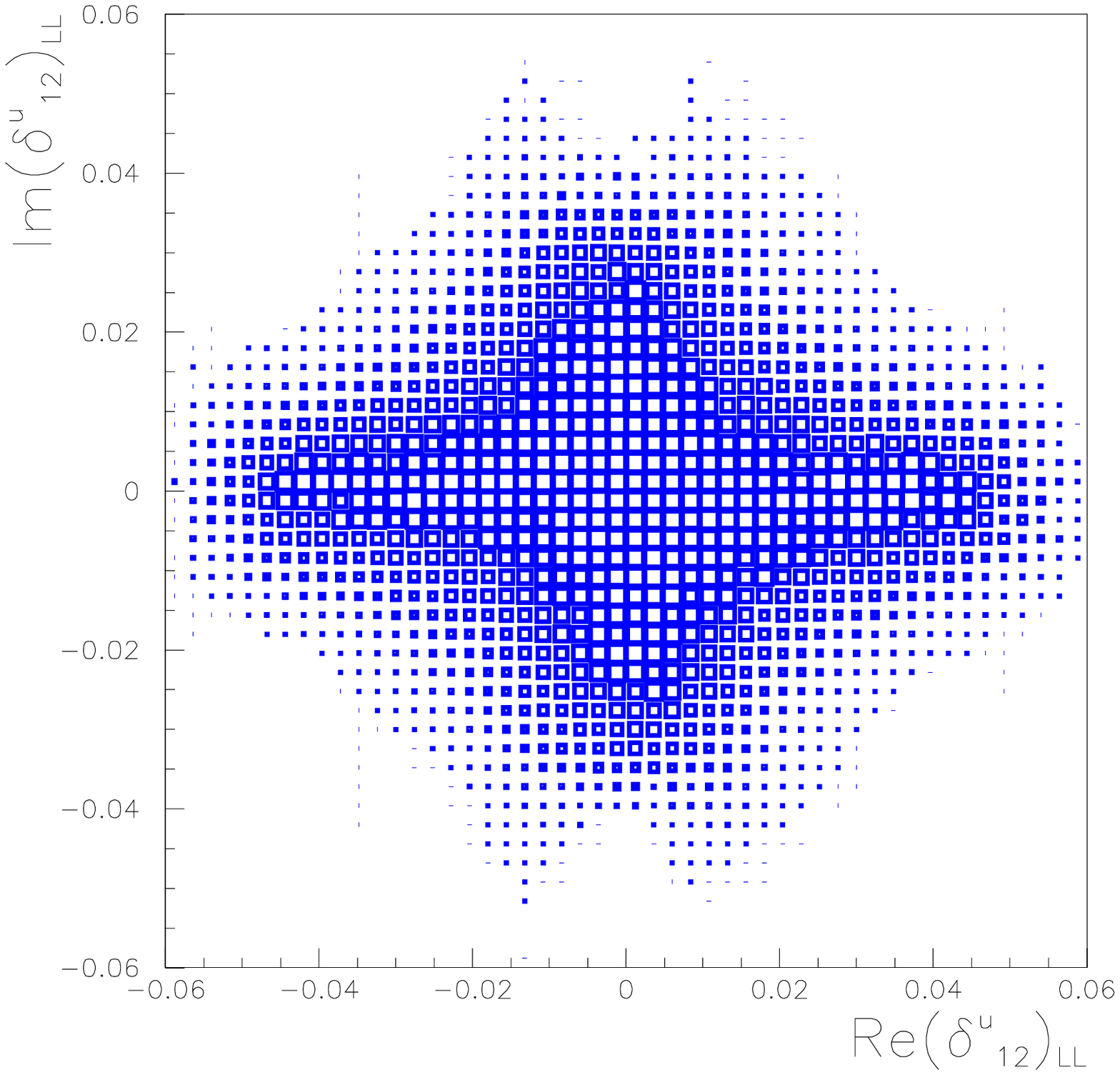}
\caption{%
Left: $A_\mathrm{NP}$ vs $\phi_\mathrm{NP}$ probability density function 
of the combined fit from Tab.~\protect\ref{tab:exp+fit} (left). 
Right: Selected region for the mass insertion $(\delta^{u}_{12})_{LL}$, 
assuming $m_{\tilde q} = m_{\tilde g} = 350$ GeV. See \cite{DDbar}.}
\label{fig:ANP}
\end{center}
\vspace{-0.5cm}
\end{figure}

A collection of (only) the most recent experimental progress on $D^0 - \bar D^0$ mixing can 
be found in Table \ref{tab:exp+fit},
\begin{table}[!tb]
\vspace{-0.5cm}
\scriptsize
\begin{center}
\begin{tabular}{|lcr|}
\hline
Parameter & Value & Ref. \\
\hline
$x^{\prime 2}_+$ & $(-0.24 \pm 0.43 \pm 0.30)\cdot 10^{-3}$ &
\cite{babarDD} \\
$x^{\prime 2}_-$ & $(-0.20 \pm 0.41 \pm 0.29)\cdot 10^{-3}$ &
\cite{babarDD} \\
$y^\prime_+$ & $(9.8 \pm 6.4 \pm 4.5)\cdot 10^{-3}$ & \cite{babarDD} \\
$y^\prime_-$ & $(9.6 \pm 6.1 \pm 4.3)\cdot 10^{-3}$ & \cite{babarDD} \\
$x$ & $(8.1 \pm 3.5)\cdot 10^{-3}$ & \cite{bellexy} \\
$y$ & $(3.7 \pm 2.9)\cdot 10^{-3}$ & \cite{bellexy} \\
$\phi$ & $(-14 \pm 19)^\circ$ & \cite{bellexy} \\
$|q/p|$ & $0.86 \pm 0.32$ & \cite{bellexy} \\
$y_\mathrm{CP}$ & $(13.1 \pm 3.2 \pm 2.5)\cdot 10^{-3}$ & \cite{belleycp} \\
$A_\Gamma$ & $(0.1 \pm 3.0 \pm 1.5)\cdot 10^{-3}$ & \cite{belleycp} \\
$\cos\delta_{K\pi}$ & $1.09\pm 0.66$ & \cite{cleo} \\
$\tau_D$ & $(0.4101 \pm 0.0015)$ ps & \cite{pdg} \\
\hline
\end{tabular}
\begin{tabular}{|ccc|}
\hline
Parameter & 68\% prob. & 95\% prob. \\
\hline
$x$ & $(6.2 \pm 2.0)\cdot 10^{-3}$ & $[0.0022,0.0105]$ \\
$y$ & $(5.5 \pm 1.4)\cdot 10^{-3}$ & $[0.0027,0.0084]$ \\
$\delta_{K\pi}$ & $(-31 \pm 39)^\circ$ & $[-103^\circ,28^\circ]$ \\
$\phi$ & $(1 \pm 7)^\circ$ & $[-15^\circ,17^\circ]$ \\
$|\frac{q}{p}|-1$ & $-0.02 \pm 0.11$ & $[-0.27,0.25]$ \\
$\vert M_{12} \vert$ & $(7.7 \pm 2.4)\cdot 10^{-3}$ ps$^{-1}$ & $[0.0030,0.0127]$ ps$^{-1}$
\\
$\Phi_{12}$ & $(2 \pm 14)^\circ\cup(179 \pm 14)^\circ$ &
$[-30^\circ,36^\circ]\cup[144^\circ,210^\circ]$ \\
$\vert \Gamma_{12} \vert$ & $(13.6 \pm 3.5)\cdot 10^{-3}$ ps$^{-1}$ & $[0.0068,0.0207]$ ps$^{-1}$\\
\hline
\end{tabular}
\end{center}
\normalsize
\caption{Left: Recent measurements related to $D^0 - \bar D^0$ mixing. 
Right: Global fit to the mixing parameters. See Ref. \cite{DDbar}.}
\label{tab:exp+fit}
\vspace{-0.5cm}
\end{table}
where $y_{CP} = \frac{\tau(D^0 \to K^- \pi^+)}{\tau(D^0 \to f_{CP})}$, $A_{\Gamma}$ is the
CP asymmetry in $D^0 \to K K$, $\delta_{K \pi}$ is the relative strong phase between wrong
sign and right sign $K \pi$ decays and $x_{\pm}, y_{\pm}$ are related to $x,y$ by a
rotation through the phase $\delta_{K\pi}$ and a subsequent one through the phase $\phi$
(detailed formulae can be found in \cite{babarDD}-\cite{DDbar}). The relevant point here is
that all the quantities listed in Table \ref{tab:exp+fit} (left) can be expressed in terms 
of $x, y, \delta_{K\pi}, \phi$ and $|q/p|$ \cite{BergmannRaz}, from which one calculates 
the fundamental mixing parameters through ($\delta = |p|^2 - |q|^2$)
\beqn
|M_{12}| \tau_D = \sqrt{\frac{x^2 + \delta^2 y^2}{4(1-\delta^2)}},~
|\Gamma_{12}| \tau_D = \sqrt{\frac{y^2 + \delta^2 x^2}{1-\delta^2}},~
\sin \Phi_{12} = 
\frac{|\Gamma_{12}|^2+4|M_{12}|^2-(x^2+y^2)|q/p|^2/\tau_D^2}{4 |M_{12} \Gamma_{12}|}.
\eeqn
The determination of $|M_{12}|, \Gamma_{12}$ and $\Phi_{12}$ can proceed through a global 
fit \cite{DDbar}, reported in Table \ref{tab:exp+fit} (right). In particular, the $M_{12}$ 
determination can then be used to place constraints on any extension of the SM. To this end, 
one can parameterize $M_{12} = |M_{12}| \exp(-i \Phi_{12}) = 
(A_{\rm SM} + A_{\rm NP} \exp(i \Phi_{\rm NP}))/\tau_D$
with the SM part, real, assumed to be flatly distributed in the range $A_{\rm SM}/\tau_D \in
[-0.015,0.015]/$ps, and obtain the implied distribution on $(A_{\rm NP},\Phi_{\rm NP})$. 
The latter, displayed in Fig. \ref{fig:ANP} (left), shows how the lack of knowledge of the 
SM contribution largely dilutes the information on the NP contribution, especially if the 
NP phase is aligned (or antialigned) with the SM (null) one.

It is clear that the information on $(A_{\rm NP},\Phi_{\rm NP})$ constrains effectively
only NP models producing in general large effects, as is the case for the MSSM with
generic flavor violation. In this instance, one can assume gluino dominance and use the
results of \cite{DF2calc} to place constraints on the mass-insertions (normalized 
off-diagonal entries of the up-squark mass matrix) $(\delta^u_{12})_{AB}$, with $AB$ the four possible
chiralities. In Fig. \ref{fig:ANP} (right) the case $AB=LL$ is reported \cite{DDbar}. The latter has
interesting consequences for models with quark-squark alignment, which tend to predict
$(\delta^u_{12})_{LL} \sim 0.2$ \cite{nirraz}. Since the bound implied by Fig. \ref{fig:ANP} 
(right) is $|(\delta^u_{12})_{LL}| = 0.037$ (95\% prob.), squark and gluino masses need to be 
raised above $\sim 2$ TeV, probably beyond the LHC reach.

\section*{Acknowledgments}
It is a pleasure to thank the organizers of EPS-HEP 2007, in particular the conveners of
the Flavor Session, for the stimulating atmosphere of the conference. I warmly thank M.
Pierini and L. Silvestrini for useful conversations.


\section*{References}

\begin{thebibliography}{50}
\bibitem{revs}
Very useful reviews on $D^0 - \bar D^0$ mixing can be found in:
  G.~Burdman and I.~Shipsey,
  Ann.\ Rev.\ Nucl.\ Part.\ Sci.\  {\bf 53} (2003) 431
  [arXiv:hep-ph/0310076].
  D.~Asner, on page 675-679 of the Review of Particle Physics, Phys.Lett.B592: 1,2004.
  Y.~Nir,
  arXiv:hep-ph/9911321.

\bibitem{petrov}
  A.~A.~Petrov,
  Int.\ J.\ Mod.\ Phys.\  A {\bf 21} (2006) 5686
  [arXiv:hep-ph/0611361].

\bibitem{ball}
  P.~Ball,
  arXiv:0704.0786 [hep-ph].

\bibitem{babarDD}
  B.~Aubert {\it et al.} [BABAR Collaboration],
  arXiv:hep-ex/0703020.

\bibitem{bellexy} 
  K.~Abe {\it et al.}  [BELLE Collaboration],
  arXiv:0704.1000 [hep-ex].

\bibitem{belleycp}
  M.~Staric {\it et al.}  [Belle Collaboration],
  arXiv:hep-ex/0703036.

\bibitem{cleo}
  D.~M.~Asner {\it et al.} [CLEO Collaboration],
  Int.\ J.\ Mod.\ Phys.\  A {\bf 21}, 5456 (2006)
  [arXiv:hep-ex/0607078].

\bibitem{pdg}
  W.~M.~Yao {\it et al.}  [Particle Data Group],
  J.\ Phys.\ G {\bf 33}, 1 (2006).

\bibitem{DDbar}
  M.~Ciuchini, E.~Franco, D.~Guadagnoli, V.~Lubicz, M.~Pierini, V.~Porretti and L.~Silvestrini,
  arXiv:hep-ph/0703204.

\bibitem{BergmannRaz}
  S.~Bergmann, Y.~Grossman, Z.~Ligeti, Y.~Nir and A.~A.~Petrov,
  Phys.\ Lett.\  B {\bf 486}, 418 (2000)
  [arXiv:hep-ph/0005181].
  G.~Raz, Phys.\ Rev.\  D {\bf 66} (2002) 057502
  [arXiv:hep-ph/0205113].

\bibitem{DF2calc}
  M.~Ciuchini, E.~Franco, D.~Guadagnoli, V.~Lubicz, V.~Porretti and
  L.~Silvestrini, 
  JHEP {\bf 0609} (2006) 013
  [arXiv:hep-ph/0606197].
  M.~Ciuchini, E.~Franco, V.~Lubicz, G.~Martinelli,
  I.~Scimemi and L.~Silvestrini,
  Nucl.\ Phys.\  B {\bf 523}, 501 (1998)
  [arXiv:hep-ph/9711402];
  A.~J.~Buras, M.~Misiak and J.~Urban,
  Nucl.\ Phys.\  B {\bf 586}, 397 (2000)
  [arXiv:hep-ph/0005183].
  D.~Becirevic, V.~Gimenez, G.~Martinelli, M.~Papinutto and J.~Reyes,
  JHEP {\bf 0204}, 025 (2002)
  [arXiv:hep-lat/0110091].

\bibitem{nirraz}
  Y.~Nir and G.~Raz,
  Phys.\ Rev.\  D {\bf 66}, 035007 (2002)
  [arXiv:hep-ph/0206064].

\end{thebibliography}
\end{document}